\documentclass {aa}
\usepackage{epsfig}
\usepackage{natbib}    
\usepackage{varioref}   
\usepackage{journals} 
\begin{document}

\thesaurus{  }
 
\title{On the star formation history of IZw~18 \thanks{This research has made use of NASA's Astrophysics Data System Abstract Service.}}

\author{F. Legrand \inst{1,2}}
\offprints{F. Legrand, legrand@iap.fr}

\institute{
  Institut d'Astrophysique de Paris, CNRS, 98bis boulevard Arago, F-75014 
  Paris, France.
  \and
  Instituto Nacional de Astrofisica, Optica y Electronica,
  Tonantzintla, Apartado Postal 51 y 216, 72000 Puebla, MEXICO
}

\date{received 27/09/1999 ; accepted 12/01/1999}

\maketitle
\markboth{ F. Legrand: On the star formation history of IZw~18}{  }
\begin{abstract}

It has been suggested that a continuous low star formation rate has
been the dominant regime in IZw~18 and in dwarf galaxies for the
lifetime of these objects \citep{LKRMHW99}. Here, we discuss and model
various star-forming histories for IZw~18. Particularly, we show that
if the metallicity observed in IZw~18 results from starburst events
only, the observed colors constrain the fraction of the metals ejected
from the galaxy to be less than 50-70\%.  We demonstrate that the
continuous star formation scenario reproduces the observed parameters
of IZw~18. A continuous star formation rate (SFR) of about $10^{-4}\
M_{\odot}\,yr^{-1}$ during 14~Gyr reproduces precisely the observed
abundances.  This SFR is comparable with the lowest SFR observed in
low surface brightness galaxies \citep{VZHSB97}.  Generalized to all
galaxies, the low continuous SFR scenario accounts for various facts:
the presence of star formation in quiescent dwarfs and LSBG, the
metallicity increase with time in the most underabundant DLA systems,
and the metal content extrapolations to the outskirts of spiral
galaxies. Also the apparent absence of galaxies with a metallicity
lower than IZw~18, the apparent absence of HI clouds without optical
counterparts, and the homogeneity of abundances in dwarfs galaxies are
natural outcomes of the scenario. This implies that, even if
starbursts are strong and important events in the life of galaxies,
their more subdued but continuous star formation regime cannot be
ignored when accounting for their chemical evolution.

\end{abstract}
 
% +++++++++++++++++++++++++++++++++++++++++++++++++++++++++++++++++++++++

\section{ Introduction }

% +++++++++++++++++++++++++++++++++++++++++++++++++++++++++++++++++++++++

A challenge of modern astrophysics is the understanding of galaxies
formation and evolution. In this exploration, low-mass dwarfs and
irregular galaxies have progressively reached a particular
place. Indeed, in hierarchical clustering theories these galaxies are
the building blocks of larger systems by merging
\citep{KWG93,PWKO96,LKGGPFVIG97}. Moreover, as primeval galaxies may
undergo rapid and strong star formation events \citep{PP67}, nearby
dwarf starburst galaxies or Blue Compact Galaxies (BCDG) of low
metallicity can also be considered as their local
counterparts. Therefore the study of low redshift starbursts is of
major interest for our understanding of galaxies formation and
evolution.

As BCDG presently undergo a strong star formation (which cannot be
maintained during a long time), but generally present a low
metallicity indicating a low level of evolution, \cite{SS72} have
proposed that these systems are young in the sense that they are
forming stars for the first time. An alternative is that they have
formed stars during strong starburst events separated by long
quiescent periods. However, most dwarf starburst galaxies show an old
underlying population indicating that they have also formed stars in
the past \citep{T83,D98phd}. Thus they are not ``young''.

Among starbursts, IZw~18, as the lowest metallicity galaxy known
locally, could be considered as the best candidate for a
truly ``young'' galaxy. However, recent studies have shown  that
even this object is not forming stars for the first time. 
Color magnitude diagrams have revealed the presence of stars
older than 1~Gyr \citep{ATG99}. \cite{LKRMHW99} have shown that
the extreme homogeneity of abundances throughout the galaxy 
\citep[see also ][]{VZWH98} cannot be explained by the metals ejected
from the massive stars formed in the current burst \citep[see also
][]{TT96}, thus indicating previous star formation. Then we need
to constrain this previous star formation and specify its nature.

It is generally accepted that the enrichment of the ISM arises by
burst phases. In the case of IZw~18, \cite{KMM95} have shown that one
single burst with intensity comparable to the present one is
sufficient to reproduce the observed abundances. However metals
ejected by massive stars could escape the galaxy, if its total mass is
lower than $\rm 10^{8} \ M_{\odot}$ \citep{MLF98}. The metallicity is
then no longer a measure of the number of bursts. On the other hand,
if the total mass of the galaxy amounts $\rm 10^{9} \ M_{\odot}$ the
metals are likely to be retained \citep{STT98}. As the total mass of
IZw~18 is likely to lie between these values \citep{VLC87,VZWH98}, the
escape of a fraction or of the totality of the newly synthesized
metals during a burst cannot be excluded. In such a case several
bursts are needed to account for the observed metallicity, their
number depending on the fraction of metals leaving the
galaxy. Nevertheless, even if metals escape, stars are likely to
remain bound and for an increasingly larger number of bursts, the old
underlying stellar population will appear progressively redder. Thus
the number of previous bursts is limited, considering the extremely
blue color of BCDG.

Between starburst events, BCDG are likely to appear as Low Surface
Brightness Galaxies (LSBG). However, studies of the latter
\citep{VZHSB97} showed that despite their low gas density, they do not
have a zero star formation rate (SFR).  LSBG indeed present a low and
possibly continuous SFR.  This led \cite{LKRMHW99} to propose that a
continuous low SFR over a Hubble time as responsible for the observed
metallicity level in the most metal-poor objects like IZw~18.

Several studies of the past star formation history of BCDG, and
specifically IZw~18, have been carried out. Most have dealt with their
chemical evolution \citep{CM82,CCPS95,KMM95} or with their
spectrophotometric properties \citep{MHK91,LH95,SL96,CMH94,MHK98}, but
rarely with both. Moreover, solely the influence of bursts has been
studied up to now, and the low continuous SFR of inter-burst phases
has been ignored.

We used a spectrophotometric model coupled with chemical evolution in
order to constrain both the abundances and the colors of the
galaxies. We also investigated the effect of a continuous and low star
formation regime. A preliminary study \citep{LK98} showed that this
scenario is plausible. Here we present detailed calculations, results
and their implications.  The model and the observational data used are
described in section \ref{section:modelisation}. The different models,
including the investigation of mass loss effect 
and the continuous star formation rate model are presented in section 
\ref{section:results}. Consequences and generalization of the
continuous SFR hypothesis are discussed in section
\ref{section:generalization}.

% +++++++++++++++++++++++++++++++++++++++++++++++++++++++++++++++++++++++

\section{Modelling the star formation history in IZw~18} \label{section:modelisation}

In order to investigate the star formation history of
IZw~18, we used the spectrophotometric model coupled with the chemical
evolution program ``STARDUST'' described by \cite{DGS99}. 
The advantage of this model is that both the metallicity and
the spectral properties of a galaxy are monitored through time.

\subsection{The model}

The main features of the model are the following:

\begin{itemize}

\item A normalized 1 $\rm M_{\odot}$ of baryonic matter galaxy is considered.

\item The SFR and the IMF are fixed and used to evaluate at each time the
   number of stars of all masses formed.

\item The stellar lifetimes are taken into account, {\it i.e.\/,} no
   instantaneous recycling approximation is used. Metals ejected
   (C,O, Fe and the total metallicity) are calculated at each time step as
   well as the number of stars of each mass. The chemical and
   spectroscopic evolution is followed in time. 

\item A fraction of the metals produced by the massive stars ($M \geq 9
   M_{\odot}$) can be expulsed from the galaxy and do not
   contribute to the enrichment.

\item The fraction of the produced metals remaining into the galaxy is
  assumed to be immediately and uniformly mixed with the interstellar
  medium. We must keep in mind that there may be a time delay between
  their production and their visibility.

\item The newly formed stars have the metallicity of the gas at the time
   of their birth.

\item The spectrum as a function of time is computed by summing the
  number of stars multiplied by their individual spectra. The nebular
  emission is not included in the model.

\item The model uses the evolutionary tracks from the Geneva group
   \citep{SSMM92,CMMS96}. The yields are from \cite{M92} for the
   massive stars ($M \geq 9 M_{\odot}$) and from \cite{RV81} for the
   lower mass stars. All the metals produced are ejected \citep[Case A
   of ][]{M92}. 

\item The stellar output spectra is computed using the
   stellar libraries from \cite{KU92} supplemented by
   \cite{BBSW89,BBSW91} for M Giants, and \cite{B95b} for M dwarfs.

\item We used a typical IMF  described as a power law 
      in the mass range 0.1-120 $\rm M_{\odot}$.       
      \begin{equation}
      \phi(m)=a.m^{-x} 
      \end{equation}
      A constant index x of 1.35 was used \citep{S55}.  We now have
      some indications that the IMF may flatten at low masses, maybe
      below $\rm 0.3\ M_{\odot}$ \citep{E99,SC98}. As the stars in this
      range (0.1-0.3 $\rm M_{\odot}$) do not contribute significantly
      to the enrichment of the ISM, nor to the colors, this will only
      act on the normalisation of the SFR in the sense that forming
      less low mass stars will decrease the total SFR requested to
      reproduce the observed abundances. However, as the SFR quoted by
      \cite{VZHS97b,VZHSB97} and reproduced in table \ref{tab:complsb}
      are computed using a Salpeter IMF down to 0.1 $\rm M_{\odot}$ we
      used this value in the model in order to compare our results
      with these previous studies. Finally, their upper mass limit is
      100 $\rm M_{\odot}$ (against 120 $\rm M_{\odot}$ in our
      models). However, the upper mass limit of the IMF do not affect
      strongly the derivation of the SFR but can modify the
      abundances. Thus using a Salpeter IMF ranging from 0.1 $\rm
      M_{\odot}$ to 120 $\rm M_{\odot}$ appears as a good compromise
      to study abundances and compare SFR with previous studies.

\item Two regimes of star formation have been investigated: 
       \begin{itemize}
         \item A continuous star formation during which the SFR is low and
           directly proportional to the total mass of available gas.
         \item A burst of star formation during which all the stars
           are formed in a rather short time. 
       \end{itemize}

\end{itemize} 

\subsection{Comparison of the model with IZw~18}

As the model is normalized to one solar mass of gas, we had to
multiply the parameters by the mass of IZw~18 in order to compare our
results with the observations. However, this normalization appears
only in the value of the SFR and in the absolute magnitude
predictions; the colors reported are independent of the choice of the
mass of the galaxy. 

The initial mass of gas in IZw~18 must lie between $\rm 6.9\,10^{7} \
and \ 8\,10^{8} \ M_{\odot} $, which are respectively the mass of HI
and the dynamical mass measured by \cite{LV80a}.  However, if only the
main component is considered, the mass must be taken between $\rm 2.6
\ 10^{7}$ (HI) and $\rm 2.6 \ 10^{8}$ (dynamical mass) as measured by
\cite{VZWH98}.  The initial mass of gas in IZw~18, in the absence of
infall, was higher than the mass of HI because of the presence of
stars and perhaps molecular $\rm H_{2}$ \citep{LV80a}.  As dark matter
can represent a non negligible fraction of the total dynamical mass,
the mass of (baryonic) gas can be lower than the dynamical mass.  We
adopted a value of $\rm 10^{8} \ M_{\odot}$ for the initial mass of
gas in IZw~18. \\

The model produces both the abundances and the spectra. We used the
spectrum to derive the expected colors in (U-B), (B-V) and (V-K). We
have adopted for comparison with the observed abundance values
reported by \cite{GSDS97} for C and \cite{SK93} for O. Most of the
published colors for IZw~18 are relatively old \citep{H77a,T83}, and
have not been corrected for the nebular contribution. \cite{S98} has
recently measured the colors of IZw~18 and corrected for the nebular
contribution. As our model does not include the nebular contribution,
we adopted Salzer's values for comparison, {\it i.e.\/,} $\rm
(U-B)=-0.88\,\pm\,0.06$ and $\rm (B-V)=-0.03\,\pm\,0.04$; \cite{T83}
has estimated that the flux measured in the IR was mainly of stellar
in origin. We thus adopted his value for (V-K)=$\rm 0.57\pm\,0.23$. \\

Finally, the model used have been compared by \cite{DGS99} with two
similar models, {\it i.e.\/,} PEGASE \citep{FRV97} and GISSEL
\citep{BC93}, and no differences larger than 0.1 magnitude were found;
this was considered the intrinsic uncertainty of the modeling process.

\section{Results of the modelisation} \label{section:results}

\subsection{Enrichment by one previous burst} \label{subsection:oneburst}

It is generally admitted that the starburst events are the main
contributors to the enrichment of the ISM. We thus used the model to
evaluate the characteristics of a single burst to reproduce the
observed oxygen abundance in IZw~18. Taking the uncertainties into
account, we found, like previous studies \citep[for example][]{KMM95},
that the present day 
abundances can be reproduced by a single burst, previous to the
current one, with a SFR of $0.065\ M_{\odot}\,yr^{-1} $ during 20~Myr.
Moreover, the contribution of this old underlying population is
too faint to modify significantly the colors of the galaxy which are
currently dominated by the newly formed massive stars. This model can
reproduce all the observations.

However, some simple arguments can rule out this model. Indeed, we
assumed that between bursts, the SFR is equal to zero, which is
certainly wrong. We will demonstrate below that even a low but
continuous SFR between bursts, as observed in LSBG, is likely to
produce significant enrichment.  Moreover, the kinetic energy
liberated in such a burst is high \citep[about $10^{40}\ erg\,s^{-1}$
using the models of ][]{MCS98}. \cite{MLF98} have suggested that for
galaxies with masses comparable with that of IZw~18, such an energy is
likely to eject out of the galaxy all the metals formed by massive
stars. If true, it means that bursts are unlikely to enrich the ISM by
much! We thus have investigated the effect of the loss of newly
synthesized elements by galactic winds. As a less extreme hypothesis,
we assumed that only a fraction of the SN ejecta (and not the
totality) leaves the galaxy. 

\subsection{The effect of metal loss}
\label{subsection:recurent}

The possibility that the energy released by the SN could eject their
products out of the galaxy was proposed earlier by
\cite{RBD88}. However, intermediate mass stars, evolving more slowly,
will eject their metals after the SN explosion of the most massive
stars. Since the kinetic energy released, mainly in stellar winds, is
lower than for the massive stars, their metal products should be
retained. This will result in a low effective enrichment in oxygen
(main product of massive stars), but in a relatively normal enrichment
in carbon (mostly produced in intermediate mass stars). This
hypothesis of ``differential galactic wind'' seems to be necessary to
reproduce the abundance measurements in some but not all the galaxies
\citep{MMT94,T98}.

\begin{figure}
\psfig{figure=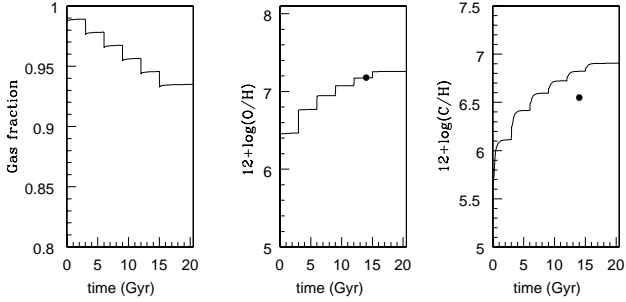,clip=,bbllx=15pt,bblly=420pt,bburx=590pt,bbury=716pt,height=4.5cm,angle=0}
\caption[]{Time evolution of the gas fraction, oxygen and carbon
  abundances for recurent bursts with a Salpeter IMF, duration of
  20~Myr and SFR=0.065~$M_{\odot}/yr$. The dots represent the measured
  abundances.} 
  \label{fig:recurentab} 
\end{figure}

\begin{figure}
\psfig{figure=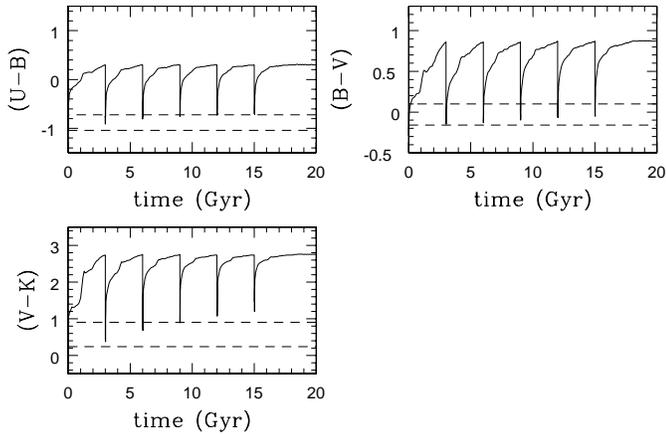 ,clip=,bbllx=28pt,bblly=330pt,bburx=570pt,bbury=700pt,height=6.cm,angle=0}
\caption[]{Colors evolution (U-B, B-V, V-K) for recurent bursts with a
  Salpeter IMF, duration of 20~Myr and SFR=0.065~$M_{\odot}/yr$. The
  plots include the additional 0.1 uncertainty (from the comparison
  between the models) with the observational errors to define a region
  (between dashed lines) of compatibility between the model and the
  observations. The solid lines represent the prediction of the
  model. If the solid line goes into the region between dashed lines,
  it means that the model does reproduce the observations within the
  error bars.}
  \label{fig:recurentco} 
\end{figure}

If a fraction of metals escapes from the galaxy during a burst, the
number of bursts necessary to reach the observed abundance in IZw~18
will be larger. We thus ran a model in which 80\% of the metals
produced by stars more massive than 9 $M_{\odot}$ left the galaxy and
did not contribute to its chemical enrichment. Assuming the same
parameters for the bursts as previously, five bursts were necessary to
reach the oxygen abundance level seen in IZw~18. We thus assumed
recurrent bursts occurring every 3~Gyr. The results of this model are
shown in figures \ref{fig:recurentab} and \ref{fig:recurentco}.

Figure \ref{fig:recurentab} shows that if the oxygen abundance is
reproduced after 5 bursts, the differential winds hypothesis results
in an overproduction of carbon. Moreover, if the carbon abundance in
IZw~18 is lower than the measurements of \cite{GSDS97}, as suggested
by \cite{IT99}, the discrepancy is even larger and completely rules
out this model. However, the uncertainties on the yields remain large
\citep[see for example][]{PR98,P99}. 
For example, the detection of WR in IZw~18 \citep{LKRMHW97} can imply
that the mass loss rate of massive stars are twice the standard ones
and the metals produced by the intermediate mass stars may also leave
the galaxy. We thus think that this argument alone is not strong
enough to invalidate definitively this scenario.

On the other hand, figure \ref{fig:recurentco} shows that after four
bursts, the expected colors (essentially V-K) do not correspond with
those observed. This is due to the fact that the old population
remaining from the previous star formation events contributes more and
more and reddens the colors. This constraint is strong, since it is
difficult to ignore old population. Of course, the constraint from the
observed colors is not really the number of previous bursts (because
it depends on their strength) but the ratio of young stars (formed in
the current burst) versus old stars (remaining from all the previous
star formation events). We thus ran another model with recurent bursts
of intensity comparable to the current one \citep{MHK98} every
1.5~Gyr. This model shows, like in figure \ref{fig:recurentco}, that the
observed colors become incompatible with observations after 6 of these
bursts. It thus appears that the total mass of stars ever formed
previously to the current burst cannot be larger than 6 times the
total mass of stars involved in the current burst. Only 2 or 3 of
these bursts produce enought metals to account for the observed
abundances (if all the metals remain into the galaxy). Thus if the
present day metallicity results from previous star formation events
{\it with mass loss}, this constrains the fraction of metals lost by
the galaxy to be lower than 50-70\%. For the same reason, this rules
out models inferring a large number of bursts in which most of the
metals leave the galaxy or produce a hot metal-rich halo as suggested
by \cite{PC87}.

\subsection{A continuous low star formation rate} \label{subsection:continuous}

As discussed by \cite{LKRMHW99}, metals observed in IZw~18, and also
in other starbursts galaxies, result from a previous star formation
episode. Assuming that the present burst in IZw~18 is the first one,
we evaluated the continuous star formation rate required to reproduce
the observed {\bf oxygen} abundance after 14~Gyr. In this scenario, a
mild star formation process would have started a long time ago but the
galaxy would presently undergo its first strong starburst event. We
found that a SFR of only $ 10^{-4}\,g\ M_{\odot}\,yr^{-1} $ ({\it
i.e.\/,} $ 10^{-3}\,g\ M_{\odot}\,Gyr^{-1}\ $ by unity of mass of gas
in the galaxy), where $g$ is the fraction of gas (in mass) available,
can reproduce the observed oxygen abundance in IZw~18. Moreover, this
model reproduces perfectly the carbon abundance measured by
\cite{GSDS97}. The kinetic energy injection rate, evaluated using the
models of \cite{MCS98}, is for this scenario of $\rm 9\,10^{36}\
erg\,s^{-1}$, {\it i.e.\/,} probably insufficient to eject the metals
out of the galaxy \citep{MLF98,STT98}.

Moreover, the kinetic energy is not deposited in one single region
like in a burst, but as the continuous star formation is supposed to
occur sporadically in location, the injected energy is diluted over
the whole galaxy, reducing the efficiency of ejection of the metals
\citep[see also ][]{SS99}. For these reasons, we assumed that all the
metals produced by the continuous star formation rate are retained in
the galaxy. Of course, this continuous star formation regime represent
an extreme case. We cannot rules out the existence of intermediate
models in which the star formation history would be a succession of
very small bursts without or with very low metal loss. However, these
models will appear, in average, as a rather continuous star formation
rate. The motivation for preferring a continuous star formation rate is
principally the absence of observational evidences for gas rich
galaxies with a SFR equal to zero, even among LSBG.

\begin{figure}
\psfig{figure=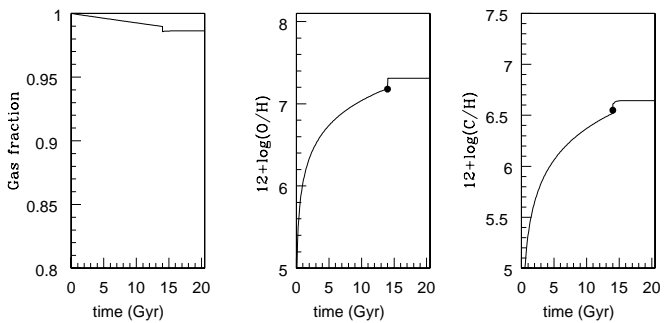,clip=,bbllx=15pt,bblly=420pt,bburx=600pt,
  bbury=700pt,height=4.5cm,angle=0}
\caption[]{Time evolution of the gas fraction, oxygen and carbon
  abundances for the continuous SFR model with a Salpeter IMF. The
  dots represent the measured abundances.}
  \label{fig:cstSFRabund} 
\end{figure}

\begin{figure}
\psfig{figure=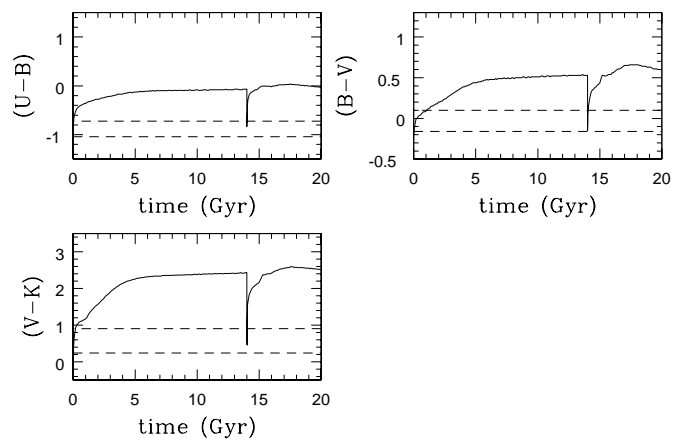,clip=,bbllx=28pt,bblly=330pt,bburx=590pt,
  bbury=700pt,height=6.cm,angle=0}
\caption[]{Colors evolution (U-B, B-V, V-K) for the continuous SFR
  model with a Salpeter IMF. The dashed lines delimit the zone of
  compatibility between the model and the observations.}
  \label{fig:cstSFRcolors} 
\end{figure}

Finally, in order to compare the colors predicted by the model, we
added the present burst at 14~Gyr. The characteristics for the current
burst, {\it i.e.\/,} a SFR of $\rm 0.023\ M_{\odot}\,yr^{-1}$ during
20~Myrs, were taken from \cite{MHK98}.  The evolution with time of the
gas fraction, oxygen and carbon abundances is presented in
Fig. \ref{fig:cstSFRabund}, whereas the evolution of the colors is
shown in Fig. \ref{fig:cstSFRcolors}.\\

Note that the observations of \cite{T83} were done using a 8''
circular aperture, which is smaller than IZw~18.  Thus the total
asymptotic magnitudes may be smaller than the ones measured by
\cite{T83}. \cite{D98phd} has shown that this difference can be as
large as 3 magnitudes, due to the presence of an old underlying
stellar population. However, as IZw~18 is a very unevolved object, the
old underlying population must be very faint. We have evaluated the
difference between the observations of \cite{T83} and the total
magnitude expected from our model in the case of the existence of an
old underlying population, due to continuous star formation, extending
uniformly over the whole galaxy. We assumed that the old underlying
population extends over 60'' $\times$ 45'' as discussed later.  If $m$
is the total magnitude emitted in a band, it can be written as 

\begin{equation}
\rm m=-2.5Log(E_{b}+E_{ci8}+E_{co8})
\end{equation}

\noindent with $\rm E_{b}$ the flux emitted by the burst 
\citep[localized in the central region included in the measurement
of][]{T83}, $\rm E_{ci8}$ and $\rm E_{co8}$ the fluxes emitted by the
old underlying stellar population inside and outside the 8'' aperture,
respectively. The magnitude measured by \cite{T83} are then:

\begin{equation}
\rm m_{8}=-2.5Log(E_{b}+E_{ci8})
\end{equation}

$\rm E_{co8}$ can be evaluated using the flux of the old underlying
stellar population predicted by the model. Under these assumptions,
the magnitude measured by \cite{T83} should be decreased by 0.20 in J
and 0.24 in K. Thus this does not change the main results and the
model predicts colors consistent with the observations. \\

The fraction of gas consumed remains very low; thus $g$ is always
close to 1. This means that the SFR is rather constant.  The model is
thus fully compatible with all the observations within the error bars.

\subsection{First consequences}

If we suppose that the continuous star formation occurs sporadically
over the whole galaxy like in LSBG \citep{VZHSB97}, the observed
homogeneity of abundances (within the NW region and also between NW
and SE regions of IZw~18) is a natural consequence of this
scenario. The uniformly distributed star formation and the long time
evolution (14~Gyr) ensure the dispersal and homogenizing of the metals
over the whole galaxy. \\

We evaluated the number of stars with a
mass greater than 8 $\rm M_{\odot}$, formed over 14~Gyr, to be around
12000. This corresponds to 120 massive stars (typically an open
cluster) formed every 140~Myr. Taking their lifetime into account, we
expect to see around 13 stars with mass greater than 8 $\rm M_{\odot}$
at a given epoch. We also evaluated the SN rate expected to be $\rm
7.5 \ 10^{-7} \ yr^{-1}$, or relatively to the mass of gas, around $\rm
10^{-14} \ yr^{-1}.M_{\odot}^{-1}$. This can be compared to the SN
rate in our galaxy which amount $\rm 10^{-13} \
yr^{-1}\,M_{\odot}^{-1}$ \citep{TLS94}. \\

\subsection{Threshold and efficiency of star formation}\label{subsection:threshold}

There is a relationship between the star formation rate and the gas
surface density \citep{S59}. A simple power law describes this link at
``high'' density, but this relationship breaks down under a critical
threshold \citep{K89,K98}. The existence of a gas density critical
threshold under which star formation would be inhibited had been
proposed by \cite{Q72}. This threshold would be associated with large
scale gravitational instabilities for formation of massive clouds. In
a thin isothermal disk, the critical gas surface density threshold
\citep{T64,C81} is \citep{K89}:

\begin{equation}
\Sigma_{c}=\alpha\frac{\kappa c }{3.36G}
\end{equation}

with

\begin{equation}
\kappa=1.41\frac{V}{R}(1+\frac{R}{V}\frac{dV}{dR})^{1/2}
\end{equation}

\noindent where $c$ is the dispersion velocity in the gas, $\kappa$ is 
the epicyclic frequency (derived from the rotation curve), and $V$ the
rotation velocity at the distance $R$ from the center of the galaxy.
However, \cite{VZHSB97} have shown that despite a gas density lower
than the threshold, LSBG are undergoing star formation.  We suggest
that in low density objects, the density can fluctuate locally and get
above the threshold in some places. This could induce localized and
faint star formation \citep{VZHSB97,S99}. For IZw~18, observations of
\cite{VZWH98} and \cite{PBCKL97} reveal a solid body rotation (in the
central part) with parameter $\frac{dV}{dR}=70 \ \rm
km\,s^{-1}\,kpc^{-1}$. Using a dispersion velocity in the gas of 12
$\rm km\,s^{-1}$ \citep{VZWH98}, the critical threshold is of the
order of $\rm 1.5\,10^{22}$ atoms\,$\rm cm^{-2}$. On the other hand,
\cite{VZWH98} have shown that the abundances in the HI halo are
comparable to the one in the HII region. This suggest that star
formation may have also occurred quite far away from the central
regions. This is reinforced by the observation of star formation in
regions with density lower than the threshold \citep{VZHSB97}. As most
of the HI gas responsible of the absorption measured by \cite{KLSV94}
is concentrated at density higher than $\rm 10^{20}$ atoms\,$\rm
cm^{-2}$, we assumed that this level represent the density limit for
the continuous star formation. This appears as a lower limit because
metal produced by star formation in more inner regions can also have
been dispersed and mixed at larger distances.  This implies that local
fluctuations in the density should be up to a factor of 150 and that
the continuous SFR could occur over a surface of $60\times45''\ $
\citep{VZWH98}. \\

The continuous star formation rate evaluated is very low. From the
work of \cite{WS89}, we can compute the star formation efficiency of
such a process. According to these authors, the star formation rate at
a distance $r$ of the center and a time $t$ is:

\begin{equation}
\psi(r,t)=\epsilon . \Omega(r) . \mu_{HI}(r,t) 
\end{equation}

\noindent where $\epsilon$ is the star formation efficiency, 
$\Omega(r)$ the local angular frequency and $\mu_{HI}(r,t)$ the
surface density in HI. Using the rotation curves computed by
\cite{VZWH98} and \cite{PBCKL97}, we found that $\Omega=\rm
(8.76\,10^{7} \ yr )^{-1}$. The HI mass measured by \cite{VZWH98}
using a surface of 2.3$\times$3 kpc gives a mean surface density of
$\rm 3 \ M_{\odot}\,pc^{-2}$. The star formation rate is then
$\psi=0.18\,\epsilon \ \rm M_{\odot}\,yr^{-1}$, which compared with the
results of our model ($\rm 10^{-4} \ M_{\odot}\,yr^{-1}$), leads to a
very low star formation efficiency of $\epsilon \sim 6\,10^{-4}$. \\

\section{Generalization of the continuous star formation
  hypothesis}\label{section:generalization} 

\subsection{Underlying population and faint objects population}

If the starburst in IZw~18 is the first one in its history, we can
expect that objects which have not yet undergone a burst (but only a
continuous star formation rate) do exist. They should then look like
IZw~18 just before the present burst. Our modeling predicts that after
14~Gyr of continuous star formation, the magnitude of a IZw~18-like
object is of the order of 20 in V and 17.5 in K. These magnitudes
represent the brightness of the old underlying stellar population in
IZw~18.  Assuming that the continuous star formation occured in
regions where the HI column density was greater than $\rm 10^{20} \
cm^{-2}$, {\it i.e.\/,} $60\times45''\ $ \citep{VZWH98}, the expected
surface brightness would be of the order of 28 $\rm mag\,arcsec^{2}$
in V and 26 $\rm mag\,arcsec^{2}$ in K. These values are an upper
limit (in $\rm mag\,arcsec^{2}$); if a fraction of metals is ejected
out of the galaxy, the SFR 
needed to produce the observed abundances will be higher and the total
luminosity and surface brightness will be increased. Moreover, as
discussed in section \ref{subsection:threshold}, the density limit
adopted for the continuous SFR is a lower limit and the region where
the continuous SFR can occur may be smaller, resulting in higher
surface brightness. However, the extreme
faintness of the old underlying population probably explains why no
strong evidence for its existence has been found in IZw~18
\citep{T83,HT95} until recently when reanalyzing HST archive images 
\cite{ATG99} found stars older than 1~Gyr.
In all the cases, these very low surface brightness levels will be
reachable with 8m class telescopes like the VLT and Gemini for rather
closeby objects.  A search for a faint old underlying population, in
the external part of dwarf galaxies like IZw~18, resulting from a
continuous star formation process, is planned. If successful, this
will support the existence of a class of very low surface brightness
galaxies, which never underwent a burst, but evolved through a
continuous weak star formation rate.\\

\subsection{LSBG as quiescent counterparts of starbursts}

We also studied the state of dwarf galaxies between bursts. IZw~18
appears as an extreme object which could present for the first time a
strong star formation event. The chemical abundance levels in most of
the starbursts galaxies suggest that they have undergone at least two
or three previous bursts. Using the model described below, we
investigated the simple following star formation history for a dwarf
galaxy with mass, size and distance comparable to IZw~18:

\begin{itemize}
\item A continuous star formation rate since 14~Gyr.
\item Two bursts with a SFR of $0.065 \ M_{\odot}\,yr^{-1}$ and a
  duration of 20~Myr, respectively at 8 and 11~Gyr.
\end{itemize}

The absolute magnitudes predicted are around -10 and the
surface brightness about 25 $\rm mag\,arcsec^{-2}$ in B. These values
correspond to objects at the extreme end of the luminosity
function of the galaxies as observed by \cite{LSY93}. 

We have also compared the continuous SFR required for IZw~18 ($
10^{-4}\ M_{\odot}\ yr^{-1} $) with those
measured by \cite{VZHS97a,VZHS97b,VZHSB97} in low surface
brightness galaxies. As these objects have different sizes and masses,
we normalized the SFR to the total HI mass. For IZw~18, the HI
mass lies between $\rm 2.6\,10^{7}\ M_{\odot}$ \citep[][for the
main body]{LV80a,VZWH98} and $\rm 6.9\,10^{7}\ M_{\odot}$ \citep[][for
the whole galaxy, including the diffuse low column density
component]{VZWH98}. The comparison is shown in table
\ref{tab:complsb}. It appears that the continuous SFR as predicted by our
scenario is comparable, relative to the HI mass, to the lowest
SFR observed in quiescent and low surface brightness
galaxies (like for example in UGC8024 or UGC9218).

We then conclude that LSBG and quiescent dwarfs are likely to be the
quiescent counterparts of starburst galaxies.

%....... Table 1Zw 18 vs LSBG..................................

    \begin{table}
      \caption {Comparison between the SFR measured in quiescent
        dwarfs and LSBG and the continuous SFR predicted for
        IZw~18.}
      \label{tab:complsb} 
      \tiny
      \begin{flushleft}
        \begin{tabular}{lcrrlrl}
          \hline
Name       &  Type$^{a}$               & Mb   & $M_{HI}$         & SFR       & $\frac{SFR}{M_{HI}}$ & $\rm ref^{b}$ \\
          &                       &  &$\rm 10^{7}\,M_{\odot}$&$\rm M_{\odot}/yr$& &  \\
\hline
UGCA20   &       Irr/LSBG   &     -14.9 &   25  &   0.0083   &  3.32e-11 &   V97c \\
UGC2684  &        dW/LSBG   &     -13.7 &   15  &   0.0015   &  1.00e-11 &   V97c \\
UGC2984  &         ?/LSBG   &     -18.4 &  500  &   0.35     &  7.00e-11 &   V97c \\
UGC3174  &       Irr/LSBG   &     -15.7 &   61  &   0.0086   &  1.41e-11 &   V97c \\
UGC5716  &       dSp/LSBG   &     -16.3 &  140  &   0.023    &  1.64e-11 &   V97c \\
UGC7178  &        dI/LSBG   &     -16.6 &  140  &   0.022    &  1.57e-11 &   V97c \\
UGC11820 &       dSp/LSBG   &     -17.7 &  440  &   0.074    &  1.68e-11 &   V97c \\
UGC191   &       dSp/qui    &     -18.2 &  320  &   0.15     &  4.69e-11 &   V97c \\
UGC300   &       dwf/LSBG   &     -16.5 &  93.3 &   0.03     &  3.26e-11 &   V97b \\
UGC521   &       Irr/LSBG   &     -15.8 &  95.5 &   0.023    &  2.41e-11 &   V97b \\
UGC634   &       dSp/qui    &     -17.7 &  470  &   0.029    &  6.17e-12 &   V97c \\
UGC891   &       dWf/qui    &     -16.3 &  100  &   0.014    &  1.40e-11 &   V97c \\
UGC1175  &        ?/qui     &     ?     &  102  &   0.0037   &  3.63e-12 &   V97b \\
UGC2162  &       dI/qui     &     ?     &  71   &   0.0059   &  8.31e-12 &   V97b \\
UGC2535  &       dI/qui     &     ?     &  275  &   <0.003   &  1.09e-12 &   V97b \\
UGC3050  &        S/qui     &     ?     &  479  &   0.13     &  2.71e-11 &   V97b \\
UGC3672  &       Irr/LSBG   &     -16.6 &  129  &   0.021    &  1.63e-11 &   V97b \\
UGC4660  &       dSp/LSBG   &     -17.8 &  363  &   0.025    &  6.89e-12 &   V97b \\
UGC4762  &       dI/qui     &     ?     &  100  &   0.0099   &  9.90e-12 &   V97b \\
UGC5764  &       dI/qui     &     -13.9 &   34  &   0.0098   &  2.88e-11 &   V97c \\
UGC5829  &       dI/qui     &     ?     &  182  &   0.080    &  4.40e-11 &   V97b \\
UGC7300  &       dI/qui     &     ?     &  129  &   0.016    &  1.24e-11 &   V97b \\
UGC8024  &       dI/qui     &     ?     &  49   &   0.0019   &  3.88e-12 &   V97b \\
UGCA357  &       dI/LSBG    &     -16.3 &  190.5&   0.034    &  1.79e-11 &   V97b \\
UGC9128  &       dI/qui     &     ?     &  3.55 &   0.00017  &  4.79e-12 &   V97b \\
Haro43   &       S /LSBG    &     -17.1 &  229  &   0.092    &  4.02e-11 &   V97b \\
UGC9762  &       dSp/LSBG   &     -18.0 &  479  &   0.05     &  1.04e-11 &   V97b \\
UGC10281 &       dwf/LSBG   &     -17.1 &  67.6 &   0.012    &  1.78e-11 &   V97b \\
\hline                                                                    
IZw~18$^{c}$    &       dI         &     -13.92&  6.9  &   0.0001   &  1.45e-12 &     \\    
IZw~18$^{c}$    &       dI         &     -13.92&  2.6  &   0.0001   &  3.85e-12 &     \\    
          \hline
        \end{tabular}
      \end{flushleft}
      \scriptsize{$ ^{a}$ Morphological type / LSBG or quiescent dwarf
        (qui) \\
        $ ^{b}$  V97b: \cite{VZHS97b}; V97c: \cite{VZHSB97} \\
        $ ^{c}$ The two values given depends if the total
        or only the main component HI mass is adopted. \\
}

    \end{table}
%....... Fin Table  ...................................

\subsection{Generalization to all galaxies}

If a continuous star formation rate exists in IZw~18, it must
exist in other dwarf galaxies, and may be, in all
galaxies. There are some hints for such a hypothesis.

For example, the extreme outerparts of spirals, where no strong star
formation event occurred and where the metals formed in the more
active inner zones have not diffused, must have low abundances and low
surface brightness. For example, let us assume that the bulk of
``strong'' star formation occurs in a spiral galaxy at distances less
than the optical radius ($\sim$10 kpc). \cite{RK95} have shown that 
metals can be dispersed at scales up to 10 kpc in about 1~Gyr, so
extending their results we can expect that if
recent (less than 1~Gyr ago) star formation occurred in external
regions located at 10 kpc from the center (one optical radius), the
newly formed metals could affect abundances at distances up to 20 kpc
form the center in few Gyrs. If this star formation is relatively
recent, we expect that the most external region of the disk (more than
2-3 optical radii) will not be affected by ``strong'' star formation
events their metallicity will be solely the result of the
``underlying'' low continuous star formation rate. Extrapolations of
metallicity gradients in spiral galaxies lead to abundances comparable
to that of IZw~18 at radial distances of about three optical radii
\citep{FWG98a,HW99}. This corresponds to the size of the halos or
disks susceptible to give rise to metallic absorption in quasar
spectra \citep{BB91}.

\begin{figure}
\psfig{figure=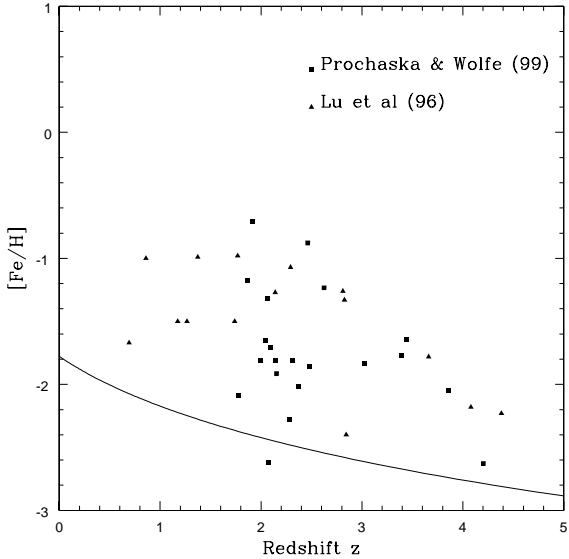 ,height=8cm,angle=0}
\caption[]{Comparison of the predicted and observed evolution with
  redshift of the abundance [Fe/H]. The points represent the data from
  \cite{LSBCV96} and \cite{PW99} and the solid line the model prediction
  for a constant star formation rate.}
  \label{fig:compdla} 
\end{figure}

We showed that a continuous low star formation rate results in a
steady increase of the metallicity of the interstellar gas. We have
compared the evolution of the iron abundance predicted from our model
with the measurements in DLA systems in Fig. \ref{fig:compdla}. The
abundances predicted by the model mimic the lower envelope of these
measurements. If we assume that these absorption systems are
associated with galaxy halos \citep{LBTW95,TLS97}, this indicates that
such a process can account for a minimal enrichment of the ISM. One
measurement appears lower than the model prediction. However, Fe atoms
are likely to condensate into grains \citep{LSB98,PKSH97}, so the iron
abundance measurements are only lower limits of the real iron
abundances. Moreover \cite{BB91} have shown that absorptions should
occur at distances of up to 4 Holmberg radius. These regions are
likely to present very low density, and may be too under-critical to
allow star formation, even in the low SFR regime described
here. Moreover, the partial ionization of these regions by the diffuse
ionizing background \citep{VG91,CSS89,MA90,CS93a,CS93b} will also
contribute to prevent star formation.  Their metallicity could thus be
due only to metals which have diffused from the inner regions. If true, we
can expect, in these regions, abundances lower than what is predicted
from the continuous SFR.

% +++++++++++++++++++++++++++++++++++++++++++++++++++++++++++++++++++++++

\section{Conclusion}

We have investigated different star formation histories for IZw~18
using a spectrophotometric model coupled with a chemical evolution
model of galaxies. We have shown that if the observed metallicity
results only from burst events with galactic winds, no more than
50-70~\% of the newly synthesized metals may have been ejected out of
the galaxy. This is because a larger metal loss rate will require to
form more stars to reach the measured abundances, hence resulting in
redder colors than what is observed, due to an overproduction of old
underlying low mass stars. Following the suggestion of
\cite{LKRMHW99}, we investigated the hypothesis of a low, but
continuous, SFR which should account alone for the observed
metallicity in IZw~18.  We have shown that the metals in IZw~18 are
likely to result from a mild continuous star formation rate which took
place indenpently from bursts. This star formation would be due to
local fluctuations in the density which exceeds sporadically the
threshold for star formation. Using a spectrophotometric model and a
chemical evolution model of galaxies, we demonstrated that a
continuous star formation rate as low as $10^{-4}\,M_{\odot}/yr$
occurring for 14~Gyrs can reproduce all the main parameters of
IZw~18. The generalization of this model to all galaxies accounts for
many observed facts, such as the presence of star formation in
quiescent dwarfs and LSBG, the increase with time of the metallicity
of the most underabundant DLA systems, the extrapolation of
metallicity in the outerparts of spiral galaxies, the lack of galaxies
with a metallicity lower than IZw~18, the apparent absence of HI
clouds without optical counterparts, and the homogeneity of abundances
in dwarfs galaxies. Moreover, we predict for IZw~18 and other
extremely unevolved galaxies, the presence of an old underlying
stellar population (resulting from this continuous star formation
process) at a surface brightness level of at least 28 $\rm
mag\,arcsec^{2}$ in V and 26 $\rm mag\,arcsec^{2}$ in K. Finally, we
have shown that the parameters for the low continuous star formation
rate are comparable to what is observed in LSBG and quiescent dwarfs,
suggesting that these objects could be the quiescent counterparts of
starburst galaxies.

% +++++++++++++++++++++++++++++++++++++++++++++++++++++++++++++++++++++++

\begin{acknowledgements}
This work is part of FL PhD Thesis.
I am indebted to Daniel Kunth for his advices, suggestions and support
during all this work. 
I thank J.~Devriendt, B.~Guiderdoni and R.~Sadat for having kindly
provided their model and spent time to explain its subtilities. I am
also grateful to J. Salzer for providing unpublished photometry of
IZw~18. I also thank J.R.~Roy, G.~Tenorio-Tagle, M.~Fioc, P.~Petitjean,
J.~Silk, R.~\&~E.~Terlevich, F.~Combes,  G.~Ostlin, J.~Lequeux,
M.~Cervi\~no, J.~Walsh and M.~Mas-Hesse for helpful suggestions and
discussions.  I also thank the anonymous referee for his remarks and
suggestions which helped to improve the manuscript. 
\end{acknowledgements}

\bibliography{/home/legrand/LATEX/BIBLIOGRAPHIE/bibliographie}

\end{document}